\documentclass[aps,10pt,superscriptaddress,showpacs,showkeys,nofootinbib]{revtex4-1}
\usepackage{graphicx}
\usepackage{bm}
\usepackage{amsmath,amssymb,amsfonts,latexsym}
\newcommand{\be}{\begin{equation}}
\newcommand{\ee}{\end{equation}}
\newcommand{\ba}{\begin{eqnarray}}
\newcommand{\ea}{\end{eqnarray}}
\newcommand{\la}{\langle}
\newcommand{\ra}{\rangle}
\newcommand{\di}{\mathrm{d}}
\begin{document}
%
%
%
\title{Spin-flavor structure of chiral-odd GPDs in the large--$\bm{N_c}$ limit}
\author{P.~Schweitzer}
\email{peter.schweitzer@phys.uconn.edu}
\affiliation{Institute for Theoretical Physics, T\"ubingen University, 
             Auf der Morgenstelle 14, 72076 T\"ubingen, Germany}
\affiliation{Department of Physics, University of Connecticut, 
	Storrs, CT 06269, USA}
\author{C.~Weiss}
\email{weiss@jlab.org}
\affiliation{Theory Center, Jefferson Lab, Newport News, VA 23606, USA}
\date{June 25, 2016}
\begin{abstract}
We study the spin-flavor structure of the nucleon's chiral-odd generalized 
parton distributions (transversity GPDs) in the large--$N_c$ limit of QCD. 
In contrast to the chiral-even case, only three combinations of the four 
chiral-odd GPDs are non-zero in the leading order of the $1/N_c$ expansion: 
$\bar E_T = E_T + 2\widetilde{H}_T, H_T$ and $\widetilde{E}_T$. The degeneracy
is explained by the absence of spin-orbit interactions correlating the transverse
momentum transfer with the transverse quark spin. It can also be deduced 
from the natural $N_c$--scaling of the quark--nucleon helicity amplitudes 
associated with the GPDs. 
In the GPD $\bar E_T$ the flavor--singlet component $u + d$ is leading in the 
$1/N_c$ expansion, while in $H_T$ and $\widetilde E_T$ 
it is the flavor--nonsinglet components $u - d$. The large--$N_c$
relations are consistent with the spin-flavor structure extracted from
hard exclusive $\pi^0$ and $\eta$ electroproduction data, if it is assumed that the
processes are mediated by twist--3 amplitudes involving the chiral-odd GPDs and
the chiral-odd pseudoscalar meson distribution amplitudes.
\end{abstract}
\maketitle
\section{Introduction}
\label{Sec-1:introduction}
Generalized parton distributions (or GPDs) have become an essential tool 
in the study of nucleon structure in QCD; see
Refs.~\cite{Goeke:2001tz,Diehl:2003ny,Belitsky:2005qn,Boffi:2007yc} for a
review.  They parametrize the nucleon matrix elements of quark and gluon
light-ray operators at non-zero momentum transfer and unify the concepts of
parton density and elastic form factor. As such they provide a comprehensive
description of the nucleon's quark and gluon single-particle structure and its
spin-flavor dependence. At twist-2 level the nucleon's quark structure is
described by 4 chiral-even (quark helicity-conserving) and 4 chiral-odd (quark
helicity-flipping) GPDs; the number corresponds to that of independent
quark-nucleon helicity amplitudes \cite{Diehl:2001pm}.  The chiral-even GPDs
reduce to the usual unpolarized and helicity--polarized quark parton
distribution functions (PDFs) in the limit of zero momentum transfer. These GPDs
appear in the collinear QCD factorization of amplitudes of hard exclusive
processes such as deeply virtual Compton scattering
\cite{Ji:1996nm,Radyushkin:1997ki,Belitsky:2001ns} and exclusive meson
production with longitudinal photon polarization \cite{Collins:1996fb} and can
be accessed experimentally in this way. The chiral-odd GPDs reduce to the quark
transversity PDFs in the limit of zero momentum transfer. Relating these GPDs to
hard exclusive processes has proved to be challenging.  The chiral-odd GPDs
decouple from single vector meson production at leading twist in all orders in
perturbative QCD due to the chirality requirements for massless fermions
\cite{Collins:1999un}.  It has been argued that the chiral-odd GPDs could be
probed in diffractive electroproduction of two mesons with large invariant mass
(rapidity gap) \cite{Ivanov:2002jj,Enberg:2006he,Beiyad:2010cxa}, but the
proposed kinematics is difficult to access and no data are presently available.

Recent theoretical work suggests that hard exclusive electroproduction of
pseudoscalar mesons ($\pi^0, \eta, \pi^+$) may be described by a hard scattering
mechanism involving the twist-2 chiral-odd nucleon GPDs and the twist-3
chiral-odd meson distribution amplitude \cite{Ahmad:2008hp,Goloskokov:2009ia,%
Goloskokov:2011rd,Goldstein:2012az}; see Ref.~\cite{Kroll:2016aop} for a
summary. A large chiral-odd distribution amplitude is induced by the dynamical
breaking of chiral symmetry in QCD, and its normalization can be determined
model-independently in terms of the chiral condensate
\cite{Goloskokov:2009ia,Goloskokov:2011rd}. While the mechanism is formally
power-suppressed and no strict factorization has been established at this level,
the pseudoscalar production amplitudes have been calculated in a modified hard
scattering approach, which implements suppression of large--size $q\bar q$
configurations in the meson through the QCD Sudakov form factor
\cite{Goloskokov:2009ia,Goloskokov:2011rd}.  The results agree well with the
$\pi^0$ and $\eta$ electroproduction data from the JLab CLAS experiment at 6 GeV
incident energy, regarding both the absolute cross sections and the dominance of
transverse photon amplitudes ($L/T$ ratio) inferred from the
azimuthal--angle--dependent response functions
\cite{Bedlinskiy:2012be,Kubarovsky:2016yaa}.  A tentative spin-flavor separation
of the chiral-odd GPDs has been performed by combining data in $\pi^0$ and
$\eta$ electroproduction using the different sensitivity of the two channels
\cite{Kubarovsky:2016yaa}. Further dedicated experiments in pseudoscalar meson
electroproduction are planned with the JLab 12 GeV Upgrade.

In order to interpret the pseudoscalar meson production data and assess the
potential of this method, it is necessary to gain more insight into the
properties of the chiral-odd GPDs from other sources. 
Contrary to the chiral-even GPDs, in the chiral-odd case
neither the zero-momentum transfer limit of the GPDs (transversity PDFs) nor the
local operator limit of the GPD (form factor of local tensor operator)
correspond to structures that are easily measurable, so that little useful
information can be obtained in this way. The transversity PDFs can be extracted
from polarization observables in semi-inclusive deep-inelastic scattering,
and in principle also from dilepton production in polarized proton-proton 
collisions, but the methods have large theoretical 
and experimental uncertainties; see
Refs.~\cite{Bacchetta:2012ty,Anselmino:2013vqa,Kang:2014zza} and references
therein. The form factors of local tensor operators, which constrain the lowest
$x$--moment of the chiral-odd GPDs, have been calculated in lattice QCD
\cite{Gockeler:2006zu} and in various dynamical models of nucleon structure;
see Refs.~\cite{Barone:2001sp,Burkardt:2015qoa} for a review.
The $x$--dependent chiral-odd GPDs have been studied in quark bound--state 
models of nucleon structure
\cite{Pasquini:2005dk,Burkardt:2007xm,Chakrabarti:2008mw,
Kumar:2015yta,Chakrabarti:2015ama}.  
Besides these
estimates not much is known about the properties of the chiral-odd GPDs.

The limit of a large number of colors in QCD (large--$N_c$ limit) provides 
a powerful model-independent method for studying the spin-flavor structure 
of nucleon matrix elements \cite{'tHooft:1973jz,Witten:1979kh,Coleman:1980mx}. 
The conceptual basis and practical implementation of this approach have been 
described extensively in the literature, see Ref.~\cite{Jenkins:1998wy} for 
a review. In the large--$N_c$ limit QCD becomes semi-classical, and baryons 
can be described by mean field solutions in terms of meson fields 
\cite{Witten:1979kh}. While the dynamics remains complex and cannot be solved 
exactly, and the form of the mean field solution is not known, qualitative 
insights can be obtained by exploiting known symmetry properties of the mean 
field \cite{Witten:1983tx,Balachandran:1982cb}. 
The resulting scaling relations for baryon mass splittings, meson-baryon 
coupling constants, electromagnetic and axial form factors, and other 
observables are generally in good agreement with observations 
\cite{Jenkins:1998wy,Dashen:1993jt,Jenkins:1993zu,Jenkins:1995td,Dashen:1993as}.
In matrix elements of quark bilinear operators (vector or axial vector 
currents, tensor operators) the large--$N_c$ limit identifies leading and 
subleading spin--flavor components and implies a parametric hierarchy in 
nucleon structure. The approach can be extended to parton densities 
\cite{Diakonov:1996sr,Efremov:2000ar}, where it suggests a large flavor 
asymmetry of the polarized antiquark distribution 
$\Delta\bar u - \Delta\bar d$, as supported by the recent RHIC $W^\mp$ 
production data \cite{Aggarwal:2010vc,Adare:2010xa}. The $N_c$--scaling of
chiral--odd quark distributions (transversity PDFs) was considered in 
Refs.~\cite{Pobylitsa:1996rs,Schweitzer:2001sr,Pobylitsa:2003ty},
and that of local chiral-odd operators (tensor charges) in 
Refs.~\cite{Kim:1995bq,Ledwig:2010tu}, in the context
of calculations in the chiral-quark soliton model of the large--$N_c$ nucleon.
A general method for the $1/N_c$ expansion of GPDs was described 
in Ref.~\cite{Goeke:2001tz} and applied to chiral-even GPDs. 

In this article we study the spin-flavor structure of the nucleon's 
chiral--odd GPDs in the large--$N_c$ limit and discuss its implications. 
We derive the $N_c$--scaling of the chiral--odd GPDs using the method of 
Ref.~\cite{Goeke:2001tz} and observe interesting differences between the 
chiral--even and chiral--odd cases. We show that the findings can be 
explained as the result of natural $N_c$--scaling of the nucleon--quark 
helicity amplitudes associated with the chiral-odd GPDs \cite{Diehl:2001pm,Diehl:2000xz}.
The spin-flavor structure obtained in the large--$N_c$ limit is found
to be consistent with that observed in an analysis of the JLab CLAS $\pi^0$ and $\eta$ 
hard exclusive electroproduction data, assuming that these processes are 
mediated by twist--3 chiral-odd meson distribution amplitudes.

The present study generalizes previous results in the $1/N_c$ expansion of 
chiral-even GPDs \cite{Goeke:2001tz}, quark transversity distributions 
\cite{Pobylitsa:1996rs,Schweitzer:2001sr,Pobylitsa:2003ty}, and matrix 
elements of local chiral-odd operators \cite{Kim:1995bq} and uses the formal 
apparatus developed in these earlier works. The description of the apparatus 
and explicit quotation of chiral-even results is intended only to make the 
present article readable. An intuitive and independent derivation of the 
$N_c$--scaling of the chiral-odd GPDs based on nucleon--quark helicity amplitudes
was given in Ref.~\cite{Schweitzer:2016vkq}.
\section{Chiral-odd GPDs}
\label{Sec-2:GPDs}
GPDs parametrize the non-forward nucleon matrix elements 
of QCD light--ray operators of the general form 
\cite{Goeke:2001tz,Diehl:2003ny,Belitsky:2005qn,Boffi:2007yc}
\be\label{Eq:generic-matrix-element}
	{\cal M } (\Gamma) = 
	P^+\int\frac{\di z^-}{2\pi}\,e^{ixP^+z^-}  
	\la N, p^\prime |\bar{\psi} (-z/2)\;\Gamma\,
	\psi (z/2)\,|N, p \ra \, |_{z^+=0,\;\bm{z}_T=0} ,
\ee
where $P \equiv \frac{1}{2}(p' + p)$ is the average nucleon 4--momentum, 
$z$ is a light-like distance, and the 4--vectors are described by their 
light-cone components 
$z^\pm = (z^0 \pm z^3)/\sqrt{2}, \, \bm{z}_T=(z^1,z^2),$ etc. The light-ray 
operator generally contains a gauge link along the light-like path defined by 
$z$, which we do not indicate for brevity.
$\Gamma$ denotes a generic matrix in spinor indices and defines the spin 
structure of the operator. In the chiral-even case 
the relevant spinor matrices are $\Gamma=\gamma^+$ and $\gamma^+\gamma_5$, 
and the matrix elements are parametrized as
\ba\label{Eq:def-chiral-even-GPD-1}
	{\cal M}(\gamma^+)
	&=&
	\bar{u}' \biggl[
	\gamma^+\;H 
	+\frac{i\sigma^{+j}\Delta_j}{2M_N}\;E\biggr] u \,,\\
\label{Eq:def-chiral-even-GPD-2}
	{\cal M}(\gamma^+\gamma_5)
	&=&
	\bar{u}^\prime \biggl[
	\gamma^+\gamma_5\;\widetilde{H} 
	+\frac{\gamma_5\Delta^+}{2M_N}\widetilde{E}\biggr] u .
\ea
In the chiral-odd case the spinor matrix is $\Gamma = i\sigma^{+j}\, (j = 1, 2)$,
and the matrix element is parametrized as \cite{Diehl:2001pm}
\ba\label{Eq:def-chiral-odd-GPD}
	{\cal M}(i\sigma^{+j})
	&=&
	\bar{u}\biggl[
	i\sigma^{+j}\;H_T
	+\frac{P^+\Delta^j-\Delta^+P^j}{M_N^2}\;\widetilde{H}_T\nonumber\\
	&& 
	+\frac{\gamma^+\Delta^j-\Delta^+\gamma^j}{2M_N}\;E_T 
	+\frac{\gamma^+P^j-P^+\gamma^j}{M_N}\;\widetilde{E}_T\biggr]
	u .
\ea
Here $u \equiv u (p, \lambda)$ and $u' \equiv u(p', \lambda')$ are the
bispinors of the initial and final nucleon (the choice of polarization states
will be specified later) and $\Delta \equiv p' - p$ is the 4--momentum transfer.
The GPDs $H=H(x,\xi,t)$, etc., are functions of the average 
quark plus momentum fraction $x$, the plus momentum transfer to the quark, 
$\xi = -\Delta^+/(2 P^+) = (p-p^\prime)^+/(p+p^\prime)^+$, and 
the invariant momentum transfer  to the nucleon, $t = \Delta^2$. 
For brevity we do not indicate the dependence of the GPDs on the normalization 
scale of the QCD operator. 

The quark fields in Eq.~(\ref{Eq:generic-matrix-element}) carry flavor indices (suppressed
for brevity), and the correlator is generally a matrix in flavor space. The usual 
flavor-diagonal GPDs are obtained with the operator
\be
H^f \;\; \leftrightarrow \;\; \bar\psi_f \ldots \psi_f
\hspace{2em} (f = u, d) ,
\label{flavor}
\ee
where the matrix element refers to the proton state. Alternatively one may consider the
isoscalar and isovector combinations $u\pm d$ of operators and GPDs. In the following 
we shall specify the flavor and isospin structure of the matrix element as needed.

The chirally-even and odd GPDs satisfy certain symmetry relations in $\xi$, 
resulting from time reversal invariance,
\be\label{Eq:xi-dependence}
	{\rm GPD}(x,-\xi,t) = \begin{cases}
     +\,{\rm GPD}(x,\xi,t)&{\rm for}\;\;{\rm GPD}=
	H,\; \widetilde{H},\; E,\;\widetilde{E},\;
	H_T,\;\widetilde{H}_T,\;E_T ,\\
     -\,{\rm GPD}(x,\xi,t)&{\rm for}\;\;{\rm GPD}=\widetilde{E}_T .\end{cases}
\ee 
Integration over the variable $x$ reduces the light-ray operators in 
Eq.~(\ref{Eq:generic-matrix-element}) to local operators. In the chiral-even 
case these are the vector and axial vector currents, so that the $x$--integral 
(or first moments) of the GPDs coincide with the electromagnetic and axial 
form factors of the nucleon \cite{Ji:1996nm}. In the chiral-odd case the 
local operator is the tensor operator $\bar{\psi}(0) i\sigma^{\mu\nu}\psi (0)$, 
and the first moments of the GPDs are
\be
\label{Eq:form-factors}
\int_{-1}^1\di x\, \{ H_T, \; \widetilde{H}_T, \; E_T, \; \widetilde{E}_T \}
(x,\xi,t)
\;\; = \;\; 
\{ H_T(t), \; \widetilde{H}_T(t), \; E_T(t), \; 0 \} ,
\ee
where $H_T(t), \widetilde{H}_T (t)$ and $E_T (t)$ are the nucleon's tensor 
form factors [with the same flavor structure as the GPDs, cf.~Eq.~(\ref{flavor})].
The vanishing of the first moment of $\widetilde{E}_T$ is a consequence of 
the antisymmetry in $\xi$, Eq.~(\ref{Eq:xi-dependence}); its higher moments 
are non-zero. The higher $x$--moments of chirally-even and odd GPDs are 
polynomials in $\xi$ (generalized form factors). 

In the limit of zero momentum transfer (forward limit) the chiral-even GPDs 
$H$ and $\widetilde H$ reduce, respectively, to the unpolarized and 
helicity PDFs, $H (x, \xi = 0, t = 0) = f_1$ and 
$\widetilde H (x, \xi = 0, t = 0) = g_1$. The GPDs $E_q$ and 
$\widetilde E_q$ are also non-zero in the forward limit but do not reduce to 
any known PDFs, as these GPDs correspond to nucleon helicity-flip components
of the matrix element (see Sec.~\ref{sec:helicity}). The chiral-odd GPD $H_T$ 
reduces in the forward limit to the transversity PDF,
\begin{equation}
	H_T (x, \xi = 0, t = 0) \;\; = \;\; h_1 (x) .
	\label{H_T_forward}
\end{equation}
Its first moment is known as the nucleon's tensor charge. Because the local 
tensor operator is not a conserved current, the tensor charge is 
scale--dependent and cannot directly be related to low--energy properties 
of the nucleon. The forward limit of 
the chiral-odd GPDs $\widetilde{H}_T$ and $E_T$
is not related to any known PDFs,
while $\widetilde{E}_T$ vanishes in the forward limit again 
due to its antisymmetry in $\xi$, Eq.~(\ref{Eq:xi-dependence}).
Other aspects of the GPDs, such as their partonic interpretation, 
are described in 
Refs.~\cite{Goeke:2001tz,Diehl:2003ny,Belitsky:2005qn,Boffi:2007yc}.

In Eqs.~(\ref{Eq:def-chiral-even-GPD-1})--(\ref{Eq:def-chiral-odd-GPD}) the GPDs 
appear as invariant amplitudes, arising from a particular decomposition of the matrix 
elements into bilinear forms between nucleon spinors. In applications to exclusive 
pseudoscalar meson production processes it is natural to introduce the 
combination 
\be
\bar{E}_T \; \equiv \; E_T + 2\widetilde{H}_T
\label{E_T_bar_def}
\ee
of the chiral-odd GPDs, which corresponds to a different invariant decomposition
of the matrix element Eq.~(\ref{Eq:def-chiral-odd-GPD}) \cite{Goloskokov:2011rd}. 
An alternative representation 
of the GPDs as nucleon-quark helicity amplitudes will 
be described in Sec.~\ref{sec:helicity}. 
%
%
\section{Chiral--odd GPDs in large--$N_c$ limit}
\label{Sec-3:large-Nc}
In the large--$N_c$ the nucleon mass scales as $M_N \sim N_c$, while the nucleon
size remains stable, $\sim N_c^0$. The $1/N_c$ expansion of GPDs is performed
in a class of frames where the initial and final nucleon move with 3--momenta 
$p^k, \,p^{\prime k} \sim N_c^0 \; (k = 1, 2, 3)$ and have energies 
$p^0, \, p^{\prime 0} = M_N + O(1/N_c)$, which implies an energy and momentum 
transfer $\Delta^0 \sim N_c^{-1}, \, \Delta^k \sim N_c^0$, and thus
\begin{equation}
	\Delta^i\; \sim \; N_c^0     \hspace{2em} (i = 1, 2), \hspace{2em} 
	\xi 	\; \sim \; N_c^{-1}, \hspace{2em} 
	|t| 	\; \sim \; N_c^0 . 
\end{equation}
In the partonic variable $x$ one considers the parametric region
\begin{equation}
	x \; \sim \; N_c^{-1} ,
\label{nc_scaling_x}
\end{equation}
corresponding to non-exceptional longitudinal momenta of the quarks 
and antiquarks relative to the slowly moving nucleon, 
$x M_N \sim (\textrm{nucleon size})^{-1} \sim  N_c^0$. Likewise, it is assumed that the 
normalization scale of the light-ray operator is $\sim N_c^0$, 
so that the typical quark transverse momenta are $\sim N_c^0$. 
Equation~(\ref{nc_scaling_x}) corresponds to the intuitive picture of 
a nucleon consisting of $N_c$ ``valence'' quarks and a ``sea'' of $O(N_c)$ 
quark--antiquark pairs, with each quark/antiquark carrying on average a 
fraction $\sim 1/N_c$ of the nucleon's light-cone momentum. Altogether, the 
$N_c$--scaling relations for GPDs can then be expressed in the form
\begin{equation}
	{\rm GPD}(x, \xi, t) \;\; \sim \;\; 
	N_c^k \; \times \; \textrm{function}(N_c x, N_c\xi, t) ,
	\label{nc_scaling_gpd_generic}
\end{equation}
where the scaling exponent $k$ depends on the GPD in question and the isospin 
component ($u + d, u - d$) and can be established on general grounds, 
while the scaling function on the 
right-hand-side is stable in the large--$N_c$ limit and can only be 
determined in specific dynamical models.

A practical method for performing the $1/N_c$ expansion of baryon matrix elements of 
quark bilinear operators was given in Refs.~\cite{Pobylitsa:2000tt,Goeke:2001tz},
using collective quantization of an abstract mean-field solution with known
symmetry properties. One considers a generic correlator of the form
\be
\la B^\prime, \bm{p}^\prime |\overline{\psi}_{\alpha' f'}(x')\psi_{\alpha f}(x)| 
B, \bm{p} \ra, 
\label{corr}
\ee
where the quark fields are at space-time points $x$ and $x'$, and $(\alpha, \alpha')$ 
and $(f, f')$ are the Dirac spinor and flavor indices. We restrict ourselves to the $SU(2)$ 
flavor sector and assume exact isospin symmetry. The baryon states are characterized by 
their momenta $\bm{p}$ and $\bm{p}'$, and spin--isospin quantum numbers 
$B \equiv \{S, S_3, T, T_3 \}$ and $B' \equiv \{S', S_3', T', T_3' \}$, and normalized such that 
\be\label{Eq-R:norm}
	\la B^\prime, \bm{p}^\prime| B, \bm{p} \ra
	= 2p^0(2\pi)^3\delta^{(3)}(\bm{p}^\prime - \bm{p})
	\,\delta_{B^{  }B^\prime}\;, \;\;\;
	\,\delta_{B^{  }B^\prime} \equiv 
	\delta_{S^{ }S^\prime}\,\delta_{S_3^{ }S_3^\prime} \,
	\delta_{T^{ }T^\prime}\,\delta_{T_3^{ }T_3^\prime} .
\ee
For simplicity we do not specify the color indices of the quark fields in Eq.~(\ref{corr})
and do not indicate the gauge link (in the case of GPDs the gauge link can be eliminated 
by choosing the light-cone gauge; in a more general case it can be included explicitly 
by an appropriate redefinition of the quark fields \cite{Pobylitsa:2003ty}). 
One evaluates the correlator Eq.~(\ref{corr}) starting with the expectation value of the 
bilinear operator in the localized mean field characterizing the large--$N_c$ baryon (``soliton''),
centered at the origin,
\be
\langle \overline{\psi}_{\alpha' f'}(x')\psi_{\alpha f}(x) \rangle 
\;\; = \;\; {\cal F}(x^{\prime 0}-x^0,\bm{x}',\bm{x})_{\alpha f; \alpha' f'} .
\label{vacuum}
\ee
While the specific form of the function $F$ depends on dynamics and can only be determined
in models, its symmetry properties in the large--$N_c$ limit can be established on general 
grounds. In leading order of $1/N_c$ the baryon mean field is static (time-independent), 
so that the correlator depends only on the relative time $x^{\prime 0} - x^0$. Most importantly,
the mean field intertwines spatial and isospin degrees of freedom (``hedgehog symmetry'') 
\cite{Witten:1983tx}, so that a rotation in flavor space by an $SU(2)$ matrix $R$ and a 
simultaneous spatial rotation with a rotation matrix $O(R)$ and spin rotation $S(R)$ 
leave the correlator Eq.~(\ref{vacuum}) invariant,
\be\label{Eq-R:hedgehog-symmetry}
	S(R)_{\alpha\beta}\, R_{fg}\,
	{\cal F}(x^{\prime 0}-x^0,O(R)\bm{x}',O(R)\bm{x})
        _{\beta g, \beta', g'} \,
	R^{-1}_{g'f'}\, S(R^{-1})_{\beta'\alpha'}
	\; = \; {\cal F}(x^{\prime 0}-x^0,\bm{x}',\bm{x})_{\alpha f; \alpha'f'}
 	\, ,
\ee
where
\be
O^{ji}(R) \;\; \equiv \;\; \tfrac{1}{2} \, {\rm tr} [R^{-1} \tau^j R \tau^i]
\hspace{2em} (i, j = 1, 2, 3) ,
\label{O_def}
\ee
and $O(R)\bm{x}'$ and $O(R)\bm{x}$ denote the rotation of the 3--vectors 
$\bm{x}'$ and $\bm{x}$ with the matrix $O(R)$.
The mean field breaks translational and rotational/isorotational invariance and does
not correspond to states of definite momentum and spin/isospin quantum numbers.
The matrix element between nucleon states of definite momentum and spin/isospin is
obtained by quantizing the collective motion in coordinate and isospin space and
projecting on states with appropriate quantum numbers. In this way one obtains a 
representation of the baryon matrix element Eq.~(\ref{corr}) in the form
\cite{Witten:1979kh,Pobylitsa:2000tt}
\ba
\la B^\prime, \bm{p}^\prime |\overline{\psi}_{\alpha' f'}(x')\psi_{\alpha f}(x)| 
B, \bm{p} \ra 
	&=& 2M_B N_c \; \int\di R\;\phi^\ast_{B^\prime}(R)\,\phi_{B^{ }}(R) \;
        \int\di^3X\,e^{i(\bm{p}' - \bm{p})\cdot\bm{X}} \nonumber\\
	&&\times \; R_{fg}\,
	{\cal F}(x^{\prime 0} - x^0,\bm{x}'-\bm{X},\bm{x}-\bm{X})
	_{\alpha g; \alpha' g'}
	\, (R^{-1})_{g'f'}\, + \dots ,
	\label{Eq-R:correlator-mean-field}
\ea 
where the dots indicate subleading terms in $1/N_c$ and $M_B\sim N_c$ is the baryon mass
(note that $M_{B'} = M_B$ in leading order of $1/N_c$).
The integral over the position of the center of the mean field, $\bm{X}$, with wave 
functions $\exp (-i\bm{p}\bm{X})$ and $\exp (i\bm{p}'\bm{X})$ projects the
correlator Eq.~(\ref{vacuum}) on baryon states with momenta $\bm{p}$ and $\bm{p}'$. 
The integral over the flavor rotation $R$ with rotational wave functions 
$\phi_{B^{ }}(R)$ and $\phi^\ast_{B^\prime}(R)$ projects on baryon states
with spin/isospin quantum numbers $B$ and $B'$. The hedgehog symmetry of the 
mean field [cf.~Eq.(\ref{Eq-R:hedgehog-symmetry})] implies that the baryon states
occur in representations with equal spin and isospin, $S = T$, and the rotational
wave functions are given by the Wigner finite rotation matrices as \cite{Goeke:2001tz}
\be
\phi_B (R) \; \equiv \;
\phi_{S_3T_3}^{S=T}(R) \; = \; 
\sqrt{2S+1}\;(-1)^{T_{ }+T_3} D_{-T_3,S_3}^{S=T}(R) .
\ee
Using Eq.~(\ref{Eq-R:correlator-mean-field}) one can evaluate the matrix element of
any bilinear quark operator in leading non-vanishing order of the $1/N_c$ expansion.
The hedgehog symmetry of the mean field, Eq.(\ref{Eq-R:hedgehog-symmetry}), restricts
the spin-isospin structures emerging from the rotational integral and determines the 
$N_c$--scaling of the spin-flavor components of the matrix element. These relations 
depend on the specific form of the operator considered.

The chiral even GPDs were evaluated in this way in Ref.~\cite{Goeke:2001tz}. Here
the operators are the light ray operators of Eq.~(\ref{Eq:generic-matrix-element}) 
with the chirally-even 
spinor matrices $\Gamma = \gamma^+$ and $\Gamma = \gamma^+ \gamma^5$, 
cf.~Eq.~(\ref{Eq:def-chiral-even-GPD-1}). It is instructive to perform the
integration over collective coordinates in Eq.~(\ref{Eq-R:correlator-mean-field}) 
in two steps. In the first step one considers the correlator integrated over the 
coordinate $\bm{X}$ but not yet over rotations; i.e., projected on momentum states but not yet 
on spin/isospin states. This correlator describes the GPDs of a large--$N_c$
baryon that has not yet been projected on spin-isopsin states (``soliton GPDs'').
In the chiral-even case it was found to be of the form \cite{Goeke:2001tz}
\be
M_N \int\frac{\di z^-}{2\pi}\,e^{ixP^+z^-}  
\la {\rm sol}, \bm{p}^\prime|\bar{\psi}_{f'}(-z/2)\,
\left\{ 
\begin{array}{c} 
\gamma^+ \\[3.5ex] 
\gamma^+ \gamma_5 
\end{array} \right\}
\psi_{f}(z/2)\,|{\rm sol}, \bm{p} \ra\biggl|_{z^+=0,\bm{z}_T=0}
=
\left\{
\begin{array}{l}
\displaystyle
\delta_{f'\!f} \, H_{\rm sol}
\; - \; \frac{i\varepsilon^{3jk}\Delta^j}{2M_N} \; D^k_{f'\!f} \,
E_{\rm sol}
\\[2.5ex]
\displaystyle
D^3_{f'\!f} \, \widetilde{H}_{\rm sol} 
\; - \; \frac{\Delta^3\Delta^j}{(2 M_N)^2}
\, D^j_{f'\!f} \, \widetilde{E}_{\rm sol}
\end{array} \right\} ,
\label{gpd_sol_even}
\ee
where $H_{\rm sol}, E_{\rm sol}, \widetilde{H}_{\rm sol}$ and 
$\widetilde{H}_{\rm sol}$ are functions of $x, \xi$ and $t$, and
[cf.~Eq.~(\ref{O_def})]
\ba
D^i_{f'\!f} \;\; \equiv \;\;  D^i_{f'\!f} (R) \;\; \equiv \;\; 
{\tfrac{1}{2}} (\tau^j)_{f'\!f} \; O^{ji} (R) .
\label{D_ff_def}
\ea
This expression embodies the hedgehog symmetry expressed by 
Eq.~(\ref{Eq-R:hedgehog-symmetry}): the 
flavor-singlet structure $\propto \delta_{f'\!f}$ is independent of the rotation matrix 
$R$ defining the orientation of the soliton, while the flavor-nonsinglet structures 
$\propto (\tau^j)_{f'\!f}$ are accompanied by rotation matrices and coupled with the
spatial directions defined by the light-ray operator ($z$--direction) and the momentum transfer 
$\bm{\Delta}$. In the second step one then projects the soliton matrix element 
on spin-isospin states by performing the integral over rotations, using
\be
\int\di R\;\phi_{S_3^\prime T_3^\prime}^{\ast S^\prime =T^\prime = 1/2}(R)\,
\; \phi_{S_3^{ }T_3^{ }}^{S^{ }=T^{ }=1/2}(R)
\;
\left\{ \begin{array}{c}
1
\\[2ex]
O^{ji}(R) 
\end{array}
\right\}
\;\; = \;\; 
\left\{ \begin{array}{c}
\delta_{S_3' S^{}_3} \,
\delta_{T_3' T^{}_3}
\\[2ex]
-\,\frac{1}{3} \, (\sigma^i)_{S_3^\prime S^{}_3} \, (\tau^j)_{T_3' T^{}_3} 
\end{array} \right\} .
\ee
The resulting nucleon matrix elements can be expressed in a transparent form by introducing 
a shorthand matrix notation for the nucleon spin components,
\be
\sigma^0 \; \equiv \; 
\sigma^0 (S_3', S_3) \; \equiv \; \delta_{S_3', S_3}, \hspace{2em}
\sigma^i \; \equiv \; 
\sigma^i (S_3', S_3) \; \equiv \; (\sigma^i)_{S_3', S_3} ,
\label{shorthand}
\ee
and correspondingly for the quark flavor components, 
\be
\tau^0 \; \equiv \; \tau^0 (f', f) \; \equiv \; \delta_{f'\!f}, 
\hspace{2em}
\tau^j \; \equiv \; \tau^j (f', f) \equiv (\tau^j)_{f'\!f} .
\ee
In this notation the nucleon matrix elements become (we consider the
proton with $T_3' = T_3 = 1/2$)
\ba\label{Eq:XXa}
{\cal M}(\gamma^+)
&=& \sigma^0 \, \tau^0 \, H_{\rm sol}
+ \frac{i (\bm{\Delta}\times
\bm{\sigma})^3 \, \tau^3}{3 (2M_N)}\,E_{\rm sol} ,\\
\label{Eq:XXb}
{\cal M}(\gamma^+\gamma_5)
&=& -\frac{\sigma^3 \, \tau^3}{3} \, \widetilde{H}_{\rm sol}
+ \frac{\Delta^3\, \bm{\Delta}\cdot\bm{\sigma} \, \tau^3}{3 (2M_N)^2} 
\widetilde{E}_{\rm sol} .
\ea
Equations~(\ref{Eq:XXa}) and (\ref{Eq:XXb}) express the spin--flavor symmetry 
characteristic of the large--$N_c$ limit:
the spin--singlet matrix element is also a flavor--singlet, and the spin---nonsinglet
one is a flavor--nonsinglet. They also allow one to determine the explicit $N_c$--scaling 
of the chiral--even GPDs. The $N_c$--scaling of the soliton GPDs 
in Eq.~(\ref{gpd_sol_even}) follows from the fact that the spatial size of the 
mean field is $\sim N_c^0$, and from the kinematic prefactors emerging from the
$1/N_c$--expansion of the spin structures in the matrix element,  and is given by 
[cf.~Eq.~(\ref{nc_scaling_gpd_generic})]
\be
\{ H_{\rm sol}, \, E_{\rm sol}, \, \widetilde{H}_{\rm sol}, \, \widetilde{E}_{\rm sol} \}
(x, \xi, t) \;\; \sim \;\; \{ N_c^2, \, N_c^3, \, N_c^2, \, N_c^4\} 
\; \times \; \textrm{function}(N_c x, N_c\xi, t) .
\ee
The $N_c$--scaling of the leading flavor components of the chiral--even nucleon
GPDs is thus obtained as \cite{Goeke:2001tz}
\be
\{ H^{u+d}, \, E^{u-d}, \, \widetilde{H}^{u-d}, \, \widetilde{E}^{u-d} \}
(x, \xi, t) \;\; \sim \;\; \{ N_c^2, \, N_c^3, \, N_c^2, \, N_c^4\} 
\; \times \; \textrm{function}(N_c x, N_c\xi, t) .
\label{chiral_even_leading}
\ee
The respective opposite flavor combinations are suppressed by one order in $1/N_c$, i.e.,
$H^{u-d} \sim N_c$, etc.

We now apply this method to the chiral-odd GPDs and derive their $N_c$--scaling.
The calculations are performed in complete analogy to the chiral--even case
described above \cite{Goeke:2001tz}. Using the specific decomposition of the chiral--odd
correlator Eq.~(\ref{Eq:def-chiral-odd-GPD}) 
and performing the $1/N_c$ expansion of the components,
we obtain the chiral--odd soliton GPDs as [cf.~Eqs.~(\ref{gpd_sol_even}) 
and (\ref{D_ff_def})]
\ba
&&	M_N\int\frac{\di z^-}{2\pi}\,e^{ixP^+z^-}  \nonumber
	\la {\rm sol}, \bm{p}^\prime |\bar{\psi}_{f'}(-z/2)\,i\sigma^{+j}
	\psi_{f}(z/2)\,|{\rm sol}, \bm{p}\ra\biggl|_{z^+=0,\bm{z}_T=0} \\
&&	\; = \; 
    	\frac{\Delta^j}{2M_N}
	\,\delta_{f'\!f} \,\bar{E}_{T,\,\rm sol}
	\; + \; i\varepsilon^{3jk} \,D_{f'\!f}^k \, H_{T,\,\rm sol} 
	\; + \; \frac{i\varepsilon^{jkl}\Delta^k}{2 M_N} \, D_{f'\!f}^l \,
        \widetilde{E}_{T,\,\rm sol} .
\label{gpd_sol_odd}
\ea
The hedgehog symmetry is again manifest in the structure of the right-hand side.
Notice that the large--$N_c$ matrix element has only three independent structures, 
and that the GPDs $E_T$ and $\widetilde{H}_T$ appear only in the combination 
$\bar{E}_T$, Eq.~(\ref{E_T_bar_def}).
Projecting on nucleon states ($T_3' = T_3 = 1/2$) we obtain, in the matrix notation of 
Eqs.~(\ref{Eq:XXa},~\ref{Eq:XXb}) ($\bm{e}_3$ denotes the unit vector in the 3--direction)
\ba\label{Eq:XXc}
	{\cal M}(i\sigma^{+j})
	\; &=& \; \sigma^0\, \tau^0 \,
	\frac{\Delta^j}{2M_N}\, \bar{E}_{T,\,\rm sol} 
	\; + \; \frac{(\bm{e}_3 \times\bm{\sigma})^j \, \tau^3}{3}\,H_{T,\, \rm sol}
	\; - \; \frac{(\bm{\Delta}\times\bm{\sigma})^j \, \tau^3}{3 (2 M_N)}\,
	\widetilde{E}_{T, \rm sol} \hspace{2em} (j = 1, 2).
\ea
The result again expresses the spin-flavor symmetry characteristic of the large--$N_c$
limit. The $N_c$--scaling of the chiral-odd soliton GPDs is found to be
\be
\{ \bar{E}_{T, {\rm sol}}, \, H_{T, {\rm sol}}, \, \widetilde{E}_{T, {\rm sol}} \} (x, \xi, t)
\;\; \sim \;\; \{ N_c^3, \, N_c^2, \, N_c^3  \}
\; \times \; \textrm{function}(N_c x, N_c\xi, t) .
\ee
We can thus identify the leading flavor components of the chiral-odd nucleon
GPDs and determine their $N_c$--scaling,
\be
\label{Eq:scaling-HT}
\{ \bar{E}_T^{u + d}, \, H_T^{u-d}, \, \widetilde{E}_T^{u-d} \} (x, \xi, t)
\;\; \sim \;\; \{ N_c^3, \, N_c^2, \, N_c^3 \}
\; \times \; \textrm{function}(N_c x, N_c\xi, t) .
\ee
The respective other flavor
components are suppressed by at least one power of $1/N_c$,
\be
\label{Eq:scaling-HT-subleading}
\{ \bar{E}_T^{u-d}, \, H_T^{u+d}, \, \widetilde{E}_T^{u+d} \} (x, \xi, t)
\;\; \sim \;\; \{ N_c^2, \, N_c, \, N_c^2  \}
\; \times \; \textrm{function}(N_c x, N_c\xi, t) .
\ee
These results confirm our earlier, intuitive 
derivation of the $N_c$--scaling using helicity amplitudes \cite{Schweitzer:2016vkq}.

The large--$N_c$ limit exposes an interesting difference between the chiral--even
and chiral-odd quark correlation functions in the nucleon, regarding the number 
of independent nucleon spin structure components, as described by the matrices 
$\sigma^0$ and $\sigma^i \, (i = 1, 2, 3)$. It can be exhibited by projecting the 
spin matrices $\sigma^i \, (i = 1, 2, 3)$ 
on the orthogonal 3--vectors $\bm{e}_3$ (the direction 
defined by the light-ray operator), $\bm{\Delta}_T \equiv 
\bm{\Delta} - (\bm{e}_3 \cdot \bm{\Delta}) \bm{e}_3$ (the component of $\bm{\Delta}$
orthogonal to $\bm{e}_3$), and
\be
\bm{n}_T \;\;\ \equiv \;\; \bm{e}_3 \times \bm{\Delta}
\ee
(the normal vector of the plane defined by $\bm{e}_3$ and $\bm{\Delta}_T$, or
the complement of $\bm{\Delta}_T$ in the transverse plane). In the chiral-even correlators
Eqs.~(\ref{Eq:XXa}) and (\ref{Eq:XXb}) one finds that all spin structures
\be
\sigma^0, \;\; \bm{e}_3 \cdot \bm{\sigma}, \;\; \bm{\Delta}_T \cdot \bm{\sigma}, \;\; 
\bm{n}_T \cdot \bm{\sigma}
\label{spin_structure_even}
\ee
are non-zero and occur with four independent coefficient functions. In the
chiral-odd correlators Eq.~(\ref{Eq:XXc}), however, the transverse nucleon 
spin structures $\bm{\Delta}_T \cdot \bm{\sigma}$ and 
$\bm{n}_T \cdot \bm{\sigma}$ occur only in the combination
\be
\bm{e}_3 \times \bm{\sigma} ,
\label{spin_structure_odd}
\ee
which {\em does not depend on the direction of the transverse 
momentum} $\bm{\Delta}_T$, and there are only three independent coefficient
functions. One sees that the reason why there are only 
three independent chiral-odd GPDs is that the 
large--$N_c$ nucleon does not correlate the direction of the transverse 
quark spin (as defined by the light-ray operator with $\sigma^{+j}$) with that of the
transverse nucleon spin (as contained in the spin structures $\bm{e}_3 \times \bm{\sigma}$
of the matrix element) through the nucleon's transverse momentum transfer. 
The absence of such spin-orbit interactions is specific to the leading 
order of the $1/N_c$ expansion, and we expect that higher-order corrections 
will remove the degeneracy of the transverse spin structures.
%
%
\section{Chiral--odd GPDs as helicity amplitudes}
\label{sec:helicity}
\begin{figure}
\parbox[c]{.38\textwidth}
{\includegraphics[width=.35\textwidth]{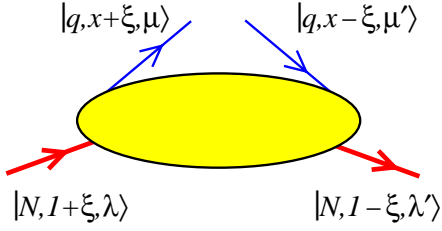}}
\hspace{.1\textwidth}
\parbox[c]{.5\textwidth}{\caption{\label{Fig-1:helicity-amplitude}
Representation of GPDs in the region $\xi < x < 1$ as nucleon--quark helicity amplitudes. 
In the nucleon and quark states (denoted as $N, q$) the second label denotes the fraction of the
light-cone plus momentum $P^+$ carried by the particle, and the third label denotes the 
light-cone helicity.}}	 
\end{figure}

Further insight into the different behavior of chiral-even and odd GPDs in the
large--$N_c$ limit can be gained by considering the representation of the GPDs as partonic
helicity amplitudes \cite{Diehl:2000xz}. This representation most naturally
appears in the region $\xi < x < 1$, where the GPDs describe the amplitude
for the ``emission'' by the nucleon of a quark with plus momentum fraction $x + \xi$ 
and subsequent ``absorption'' of a quark with $x - \xi$ 
(see Fig.~\ref{Fig-1:helicity-amplitude}). In the region $-1 < x < -\xi$ the GPDs describe 
the emission and absorption of an antiquark, while in $-\xi < x <\xi$ they describe 
the emission of a quark--antiquark pair by the nucleon. We do not need 
to consider these regions separately in the subsequent arguments.
%
%
%
%

The partonic helicity amplitudes are defined by a correlator of the form of
Eq.~(\ref{Eq:generic-matrix-element}), in which the nucleon spin states are
described as light-front helicity states and the quark spinor matrices are 
chosen
as projectors on quark light-front helicity states:
\be
	\!A_{\lambda^\prime\mu^\prime,\lambda\mu} \;\; = \;\; P^+   
	\int\frac{\di z^-}{2\pi}\,e^{ixP^+z^-}  
	\la N, p^\prime, \lambda^\prime| \, \overline{\psi}(-z/2)\; \Gamma_{\mu'\mu} \;
         \psi(z/2) \, |N, p, \lambda \ra \, |_{{z^+=0},\;{\bm{z}_T=0}\,} ,
\ee
where $\lambda (\lambda')$ are the light-front helicities of the initial (final) 
nucleon and $\mu (\mu')$ those of the initial (final) quark \cite{Diehl:2000xz}.
It is convenient to work in a reference frame where the light-cone direction is
chosen as the $z$-axis and the initial and final nucleon momenta $\bm{p}$ and 
$\bm{p^\prime}$ lie in the $x$--$z$ plane. The spinor matrices for
the light-front helicity conserving (chiral-even) and light-front 
helicity flipping (chiral-odd) amplitudes are then given by \cite{Diehl:2001pm}
\ba
\Gamma_{\pm\pm} &=& \frac{1}{4}\gamma^+(1\pm\gamma_5) ,
\\[2ex]
\Gamma_{\pm\mp} &=& \frac{i}{4}\sigma^{+1}(\pm 1 -\gamma_5)
\; = \;  \frac{i}{4}(\pm \sigma^{+1} + i \sigma^{+2}) .
\ea
Flavor components of the amplitudes can be defined in analogy to those of the 
correlator Eq.~(\ref{Eq:generic-matrix-element}) and will be specified below.
The helicity-conserving amplitudes are related to the chiral-even GPDs as
\begin{subequations}
\ba\label{Eq:amplitudes-0}
&&	A_{++,++}=\frac12\,\sqrt{1-\xi^2}
	\biggl(H+\widetilde{H}-\frac{\xi^2}{1-\xi^2}(E+\widetilde{E})\biggr) , 
	\;\;\;\;\;\;\;\;\;\\
&&	A_{-+,-+}=\frac12\,\sqrt{1-\xi^2}
	\biggl(H-\widetilde{H}-\frac{\xi^2}{1-\xi^2}(E-\widetilde{E})\biggr) , \\
&&	A_{++,-+}=\frac12\,\delta_t\;\biggl(\xi\widetilde{E}-E\biggr) , \\
&&	A_{-+,++}=\frac12\,\delta_t\;\biggl(\xi\widetilde{E}+E\biggr) ,
\label{Eq:amplitudes-0-end}\ea
\end{subequations}
while the helicity-flipping amplitudes are related to the chiral-odd GPDs as
\begin{subequations}
\ba   	\label{Eq:amplitudes-I}
&&	A_{++,+-}=\delta_t
	\biggl(\widetilde{H}_T+\frac{1-\xi}{2}(E_T+\widetilde{E}_T)\biggr) , \\
&&	A_{-+,--}=\delta_t
	\biggl(\widetilde{H}_T+\frac{1+\xi}{2}(E_T-\widetilde{E}_T)\biggr) , \\
&&	A_{++,--}=\sqrt{1-\xi^2}
	\biggl(H_T+\delta_t^2\widetilde{H}_T
	-\frac{\xi^2}{1-\xi^2}E_T
	+\frac{\xi}{1-\xi^2}\widetilde{E}_T\biggr) , \\
&&	\label{Eq:amplitudes-I-end}
 	A_{-+,+-}=\sqrt{1-\xi^2}\,
	\delta_t^2\;\widetilde{H}_T\,, \phantom{\frac11} 
\ea
\end{subequations}
where the ``kinematic'' prefactor $\delta_t$ is defined as
\be\label{Eq:delta-k}
	\delta_t = {\rm sign}\biggl(P^+\Delta^1-\Delta^+P^1\biggr)
	\frac{\sqrt{t_0-t}}{2M_N}\;,\;\;\;
	-t_0 = \frac{4M_N^2\xi^2}{1-\xi^2}\; ,
\ee
in which $-t_0$ is the minimal value of $-t$ for the given 
value of $\xi$. There are four linearly independent amplitudes in each 
sector; the other four amplitudes in each sector can be obtained from 
those in Eqs.~(\ref{Eq:amplitudes-0}--\ref{Eq:amplitudes-0-end}) and 
(\ref{Eq:amplitudes-I}--\ref{Eq:amplitudes-I-end}) by the parity 
relation \cite{Diehl:2001pm}
\be
A_{-\lambda'-\mu', -\lambda-\mu} \;\; = \;\; (-)^{\lambda' - \mu' - \lambda + \mu} \,
A_{\lambda'\mu', \lambda\mu} .
\label{parity}
\ee
Altogether, there are 8 linearly 
independent helicity amplitudes, corresponding to the total number of 
chiral-even and chiral-odd GPDs (or invariant amplitudes).

It is instructive to study the $N_c$--scaling of the partonic helicity amplitudes. 
The ``natural'' scaling of the individual helicity amplitudes for a given quark 
flavor ($f = u,d$) is 
\be\label{Eq:amplitudes-generic-scaling}
	A^f_{\lambda^\prime\mu^\prime,\lambda\mu} \;\; \sim \;\; N_c^2,
\ee
which is understood with the arguments $x, \xi$ and $t$ scaling 
as in Eq.~(\ref{nc_scaling_gpd_generic}). 
One power of $N_c$ originates from the covariant normalization
of the nucleon states in Eq.~(\ref{Eq-R:norm}), because $P^0\sim N_c$,
and another power of $N_c$ from the implicit summation over the color 
indices in the light-ray operators.
Combinations of amplitudes corresponding to definite isospin transitions
($u + d, u - d$) can vanish in leading order of the $1/N_c$ expansion due 
to the symmetries of the mean field solution (cf.~Sec.~\ref{Sec-3:large-Nc})
and have a lower scaling exponent. Using the results of Ref.~\cite{Goeke:2001tz}
and Sec.~\ref{Sec-3:large-Nc} for the $N_c$--scaling of the GPDs we can now 
identify the leading and subleading helicity amplitudes.
For the chiral-even amplitudes one obtains
\begin{subequations}
\ba\label{Eq:amplitudes-expand-0}
  	A^{u+d}_{++,++}=\tfrac12\,H^{u+d}\,, \;\;\; 
 &&	A^{u-d}_{++,++}=\tfrac12\, (\phantom{-}\widetilde{H}^{u-d}-\xi^2\widetilde{E}^{u-d}) , \\
 	A^{u+d}_{-+,-+}=\tfrac12\,H^{u+d}\,, \;\;\; 
 &&	A^{u-d}_{-+,-+}=\tfrac12\,(-\widetilde{H}^{u-d}+\xi^2\widetilde{E}^{u-d}) , \\
 	A^{u+d}_{++,-+}=0, \;\;\;\;\;\;\;\;\;\;\;\;\;\;
 &&	A^{u-d}_{++,-+}=\tfrac12\,\delta_t\; (\xi\widetilde{E}^{u-d}-E^{u-d}) , \\
 	A^{u+d}_{-+,++}=0, \;\;\;\;\;\;\;\;\;\;\;\;\;\;
 && 	A^{u-d}_{-+,++}=\tfrac12\,\delta_t\; (\xi\widetilde{E}^{u-d}+E^{u-d}) .
\label{Eq:amplitudes-expand-0-end}\ea
\end{subequations}
The expressions correspond to the leading order of the $1/N_c$ expansion,
i.e., they are accurate at the natural order ${\cal O}(N_c^2)$.
It is understood that all non-zero amplitudes receive corrections 
of order ${\cal O}(N_c)$. The amplitudes that vanish do so at order 
${\cal O}(N_c^2)$, and generically have corrections of order ${\cal O}(N_c)$. 
Notice that in the specific frame chosen here in the large--$N_c$ limit 
the kinematic factors simplify as
\be\label{Eq:xi-deltak-large-Nc}
	\delta_t = \frac{\Delta^1}{2M_N} \, ,\;\;\;
	\xi = -\,\frac{\Delta^3}{2M_N} \, .
\ee
One notices that two pairs of the chiral-even amplitudes are degenerate (up to an 
overall sign) at ${\cal O}(N_c^2)$, and two other amplitudes vanish at this order. 
The content of 
Eqs.~(\ref{Eq:amplitudes-expand-0}--\ref{Eq:amplitudes-expand-0-end}) becomes
more transparent when considering linear combinations of the chiral-even amplitudes,
\begin{subequations}
\ba\label{Eq:amplitudes-expand-1}
  	A^{u+d}_{++,++}+A^{u+d}_{-+,-+}&=&\phantom{\xi\,\delta_t\;}H^{u+d}\,, \\
 	A^{u-d}_{++,++}-A^{u-d}_{-+,-+}&=&\phantom{\xi\,\delta_t\;}\widetilde{H}^{u-d}-\xi^2\widetilde{E}^{u-d}\,, \\
 	A^{u-d}_{-+,++}+A^{u-d}_{++,-+}&=&         \xi \,\delta_t\;\widetilde{E}^{u-d}\,,\\
  	A^{u-d}_{-+,++}-A^{u-d}_{++,-+}&=&\phantom{\xi}\,\delta_t\;E^{u-d}\,.
\label{Eq:amplitudes-expand-1-end}
\ea
\end{subequations}
This representation shows that there are four independent combinations of helicity
amplitudes that appear in leading order of the $1/N_c$ expansion, which are unambiguously 
associated with  the four leading spin-flavor components of the chiral-even GPDs.
As a consequence, each of the four chiral-even GPDs has a leading flavor component:
$u + d$ in $H$, and $u- d$ in $\widetilde{H}, E,$ and $\widetilde{E}$. 

The situation is different in the case of chiral-odd helicity amplitudes.
Using the results for the $1/N_c$ expansion of Sec.~\ref{Sec-3:large-Nc} we 
obtain the following scaling behavior of the chiral-odd helicity amplitudes
at ${\mathcal O}(N_c^2)$:
\begin{subequations}
\ba
\label{Eq:amplitudes-II}
	A^{u+d}_{++,+-}=\tfrac{1}{2}\;\delta_t\; \bar E_T^{u+d} , \;\;\;
&&	A^{u-d}_{++,+-}=\phantom{-}\tfrac{1}{2}\;\delta_t\;\widetilde{E}_T^{u-d} ,\\
	A^{u+d}_{-+,--}=\tfrac{1}{2}\;\delta_t\; \bar E_T^{u+d} ,\;\;\;
&&	A^{u-d}_{-+,--}=-\tfrac{1}{2}\;\delta_t\;\widetilde{E}_T^{u-d} , \\
	A^{u+d}_{++,--}=0 \,, \;\;\; \hspace{1.3cm}
&&	A^{u-d}_{++,--}=\; H_T^{u-d}+\xi\widetilde{E}_T^{u-d} \,, \\
	A^{u+d}_{-+,+-}=0 \, , \;\;\;\hspace{1.3cm}
&&	A^{u-d}_{-+,+-}=0 .
\label{Eq:amplitudes-II-end}
\ea
\end{subequations}
Again, we obtain a more transparent representation by considering the linear
combinations
\begin{subequations}
\ba
\label{Eq:amplitudes-III}
	A^{u+d}_{++,+-}+A^{u+d}_{-+,--}&=&\delta_t\; \bar E_T^{u+d} ,\;\;\;\\
	A^{u-d}_{++,+-}-A^{u-d}_{-+,--}&=&\delta_t\;\widetilde{E}_T^{u-d} , \\
	A^{u-d}_{++,--}	&=& H_T^{u-d}+\xi\widetilde{E}_T^{u-d} \,, \\
	A^{u\pm d}_{-+,+-}	&=&0 \, .
\label{Eq:amplitudes-III-end}
\ea
\end{subequations}
One sees that in the chiral-odd case one amplitude vanishes completely:
$A_{-+,+-}=0$ for {\it both} flavor combinations $u + d$ and $u - d$.
As a result, there are only three linearly independent amplitudes that 
are non-zero in leading order of the $1/N_c$ expansion. This reflects the results 
of Sec.~\ref{Sec-3:large-Nc}, where it was found that only three independent 
GPDs are present in the large--$N_c$ nucleon, see Eqs.~(\ref{gpd_sol_odd}) et seq.
Notice that, because $A_{-+,+-}$ is related exclusively to the GPD $\widetilde{H}$
and this amplitude vanishes for any flavor combination,
it is not possible to separate the linear combination of the GPDs 
$\bar E_T = E_T + 2\widetilde{H}_T$ in leading order of the $1/N_c$ expansion.

The amplitude $A_{-+,+-}$ is unique in that it corresponds to a {\it double-helicity-flip
transition} with angular momentum exchange $\Delta J = 2$ between the active quark and 
the nucleon, i.e., the nucleon and quark helicities are flipped in opposite directions.
It is natural that for this amplitude both flavor combinations vanish in leading
order of the $1/N_c$ expansion. Because of the spin-flavor symmetry implied by the
large--$N_c$ limit the transition with $\Delta J = 2$ should be accompanied by
isospin transfer $\Delta T = 2$, which is impossible with a quark one-body operator.
This could be proved more formally by expanding the GPDs in powers of the transverse
momentum transfer, such that they can be represented by matrix elements of local 
operators (containing total derivatives) at zero transverse momentum transfer, and 
classifying the resulting local operators according to the spin-flavor symmetry
implied by the large--$N_c$ limit. The collective quantization procedure of
Sec.~\ref{Sec-3:large-Nc} \cite{Pobylitsa:2000tt,Goeke:2001tz} implements this
symmetry through the hedgehog symmetry of the mean field, Eq.~(\ref{Eq-R:hedgehog-symmetry}).

It is interesting to note that the vanishing of 
the amplitude $A_{-+,+-}$ in leading order of the $1/N_c$ expansion 
can also be derived from large--$N_c$ consistency arguments. The latter are analogous 
to the unitarity requirements imposed on meson-baryon scattering amplitudes, from which 
one can derive specific relations between meson-baryon coupling 
constants \cite{Gervais:1983wq,Dashen:1993as}. In fact, a non-vanishing 
amplitude $A^{u \pm d}_{-+,+-} \sim N_c^2$ (for any of the flavor combinations)
would imply that $\widetilde{H}^{u\pm d} \sim N_c^4$. Inserting this scaling behavior
into $A_{++,+-}$ or $A_{-+,--}$ would imply that these amplitudes should scale 
$\sim N_c^3$, which contradicts the natural scaling 
Eq.~(\ref{Eq:amplitudes-generic-scaling}).\footnote{Although 
the partonic helicity amplitudes are not strictly physical, 
they enter into the description of cross sections of certain exclusive processes
with quark helicity flip. If some of the amplitudes had a scaling $\sim N_c^3$ it is plausible 
that this would violate positivity constraints for the cross sections of some
hypothetical physical scattering processes. Whether such an argument could be
applied to chiral-odd GPDs remains an interesting question for further study.
Positivity constraints for chiral-odd GPDs were discussed in Ref.~\cite{Kirch:2005in}.}

In the discussion here we have inferred the $N_c$--scaling of the helicity amplitudes 
from that of the GPDs (or invariant amplitudes). Alternatively one may consider the 
large--$N_c$ correlators, Eqs.~(\ref{Eq:XXa}), (\ref{Eq:XXb}) and (\ref{Eq:XXc}),
directly in the particular frame $\bm{\Delta} = (\Delta^1,0,\Delta^3)$ and determine the 
helicity amplitudes from there. For reference we present in 
Appendix~\ref{app} the expressions for the correlators in that frame.
They show explicitly the degeneracy of the 
transverse spin structure of the chiral-odd correlator noted in Sec.~\ref{Sec-3:large-Nc}
[cf.~Eqs.~(\ref{spin_structure_even}) and (\ref{spin_structure_odd})], which is the
cause of the reduced number of independent chiral-odd GPDs viz.\ helicity amplitudes 
in leading order of the $1/N_c$--expansion.
%
%
\section{Flavor structure from pseudoscalar meson production data}
\label{Sec-5:comparison}
It is interesting to compare our results with preliminary data from the JLab CLAS 
exclusive pseudoscalar meson production experiments 
\cite{Bedlinskiy:2012be,Kubarovsky:2016yaa} 
(cf.\ comments in Sec.~\ref{Sec-1:introduction}).
Analysis of the azimuthal--angle dependent response functions shows that 
$|\sigma_{LT}| \ll |\sigma_{TT}|$, which indicates dominance of the twist-3 amplitudes, 
involving the chiral-odd GPDs $H_T^q$ and $\bar E_T = E_T + 2\widetilde{H}_T$,
over the twist--2 amplitudes involving the chiral-even GPD $\widetilde E^q$. A preliminary 
flavor decomposition was performed assuming dominance of the twist--3 amplitudes and
combining the data on $\pi^0$ and $\eta$ production, in which the $u$ and $d$ quark 
GPDs enter with different relative weight. Results show opposite sign of the
exclusive amplitudes $\langle H_T^u \rangle$ and $\langle H_T^d \rangle$, which is 
consistent with the leading appearance of the flavor-nonsinglet $H_T^{u - d}$ in the
$1/N_c$ expansion. (Here $\langle\ldots \rangle$ denotes the integral over $x$ of 
the GPD, weighted with the meson wave function, hard process amplitude, and Sudakov 
form factor \cite{Goloskokov:2011rd}.) The results also suggest same sign of
$\langle \bar E_T^u \rangle$ and $\langle \bar E_T^d \rangle$, which is again
consistent with the leading appearance of the flavor-singlets $E_T^{u + d}$ and 
$\widetilde H_T^{u + d}$ in the $1/N_c$ expansion. These findings should be interpreted
with several caveats: (a) the errors in the experimental extraction of
$\langle H_T^q \rangle$ and $\langle \bar E_T^q \rangle$ are substantial; 
(b) the $1/N_c$ expansion predicts only the scaling behavior, not the absolute
magnitude of the individual flavor combinations, cf.\ Eq.~(\ref{nc_scaling_gpd_generic}).

It is encouraging that the flavor structure of the amplitudes extracted from the 
$\pi^0$ and $\eta$ electroproduction data is consistent with the pattern predicted by
the $1/N_c$ expansion. Our findings further support the idea that pseudoscalar meson
production at $x_B \gtrsim 0.1$ and $Q^2 \sim \textrm{few GeV}^2$ is governed by 
the twist-3 mechanism involving the chiral-odd GPDs. 
%
%
\section{Discussion and outlook}
The large--$N_c$ limit reveals interesting characteristic differences between the
nucleon matrix elements of chiral-even and chiral-odd light-ray operators. 
While in the chiral-even case four GPDs (or invariant amplitudes) are non-zero
in the leading order of the $1/N_c$ expansion, in the chiral-odd case only
three independent GPDs appear, due to the absence of spin-orbit interactions 
correlating the transverse quark spin with the transverse momentum transfer 
to the nucleon. In the equivalent representation of GPDs as nucleon-quark 
helicity amplitudes, the same happens due to the vanishing of the double 
helicity-flip amplitude in the leading order of $1/N_c$. These conclusions are 
model-independent and do not rely on any extraneous assumptions regarding the 
internal dynamics giving rise to the partonic structure.

The leading order of the $1/N_c$ expansion predicts the scaling behavior of the leading
flavor combinations in the GPDs $\bar E_T = E_T + 2\widetilde{H}_T, H_T$ and
$\widetilde{E}_T$. Interestingly, the hard exclusive amplitudes in the twist-3 mechanism 
involve exactly these three combinations of GPDs, so that the large--$N_c$ predictions 
can be confronted with experimental observables.

The $N_c$--scaling relations of the chiral-odd GPDs described here generalize
earlier results for the $N_c$--scaling of the nucleon's transversity 
PDFs \cite{Pobylitsa:1996rs,Schweitzer:2001sr}, tensor charges \cite{Kim:1995bq}, 
and tensor form factors \cite{Ledwig:2010tu}. We note that the lattice QCD 
calculations of Ref.~\cite{Gockeler:2006zu} for the tensor form 
factors $A_{T10}(t) = H_T(t) \equiv \int dx \, H_T (x, \xi, t)$
show opposite sign for $u$ and $d$ flavors, while those for
$\bar B_{T10}(t) = \bar E_T(t) \equiv \int dx \, \bar E_T (x, \xi, t)$
show same sign for $u$ and $d$ flavors, in agreement with the leading--order 
large--$N_c$ relations Eq.~(\ref{Eq:scaling-HT}). The flavor structure of
$\bar E_T(t)$ at large $N_c$ was also studied in the bag model calculation 
of Ref.~\cite{Burkardt:2007xm} and agrees with the general result.

In the present study we have considered the leading non-vanishing order of the $1/N_c$ 
expansion of the chiral-odd nucleon matrix elements. Extension to subleading order
requires principal considerations and technical improvements. At subleading order 
the mean-field approximation to the large--$N_c$ correlation functions 
Eq.~(\ref{Eq-R:correlator-mean-field}) must include
the effects of the finite velocity of the soliton collective (iso) rotations, 
$\Omega \sim N_c^{-1}$. At the same time one must reconsider the choice of nucleon 
spinors in the invariant decomposition of the matrix elements,
Eqs.~(\ref{Eq:def-chiral-even-GPD-1}) and (\ref{Eq:def-chiral-even-GPD-2}), as the apparent 
size of ``relativistic corrections'' to a given invariant amplitude may depend on the 
choice of nucleon spinors. The choice should be guided by the symmetries of the 
leading--order approximation and incorporate corrections through a Foldy-Wouthuysen
transformation.

It would be interesting to calculate the chiral-odd GPDs in dynamical models that 
consistently implement the $N_c$--scaling, such as the chiral quark--soliton model. 
Such calculations would allow one to calculate also the scaling functions in 
the large--$N_c$ relations, Eq.~(\ref{nc_scaling_gpd_generic}), and supplement 
the scaling studies with dynamical information. $N_c$--scaling can also 
be implemented in calculations of peripheral GPDs (at impact parameters $b \sim M_\pi^{-1}$)
in chiral effective field theory \cite{Granados:2016jjl}.
%
%
\begin{acknowledgments}
In this study we greatly benefited from discussions with D.~Diakonov, 
V.~Petrov, P.~Pobylitsa, and M.~Polyakov during earlier joint work. 

This material is based upon work supported by the U.S.~Department of Energy, 
Office of Science, Office of Nuclear Physics under contract DE-AC05-06OR23177.
This work was supported by the U.S.~National Science Foundation under 
Contract No.~1406298, and by the Deutsche Forschungsgemeinschaft 
(grant VO 1049/1).
\end{acknowledgments}
\appendix
\section{Large--$N_c$ correlators in helicity frame}
\label{app}
In this appendix we express the nucleon--quark helicity amplitudes in the large--$N_c$ 
limit directly in terms of the large--$N_c$ correlators Eqs.~(\ref{Eq:XXa}), 
(\ref{Eq:XXb}) and (\ref{Eq:XXc}). To this end we consider
the correlators in the specific frame where the nucleon momenta lie in 
the $x$--$z$-plane, $\bm{\Delta} = (\Delta^1,0,\Delta^3)$, and with the 
momentum components given by Eq.~(\ref{Eq:xi-deltak-large-Nc}).
The chiral-even correlators Eqs.~(\ref{Eq:XXa},~\ref{Eq:XXb}) take the form
[in the shorthand notation of Eq.~(\ref{shorthand})]
\ba\label{Eq:YYa}
{\cal M}(\gamma^+)
&=& \sigma^0 \, \tau^0 \, H_{\rm sol}
\; + \; \frac{i \sigma^2 \, \tau^3}{3} \, \delta_t \, E_{\rm sol} ,\\
\label{Eq:YYb}
{\cal M}(\gamma^+\gamma_5)
&=& -  \frac{\sigma^1 \, \tau^3}{3}
\delta_t \xi \, \widetilde{E}_{\rm sol} \,
\; + \; \frac{\sigma^3 \, \tau^3}{3} 
(\xi^2 \widetilde{E}_{\rm sol} - \widetilde{H}_{\rm sol}) .
\ea
We now use that (a) the Dirac matrices $\gamma^+$ and $\gamma^+\gamma_5$ are the
sum and difference of the quark helicity projectors,
\be
\left.
\begin{array}{l}
\gamma^+ 
\\[1ex]
\gamma^+ \gamma_5
\end{array} 
\right\} \;\; = \;\; 2 (\Gamma_{++} \pm \Gamma_{--}) ;
\ee
(b) the nucleon light-front helicity can be identified with the ordinary spin
projection on the 3--axis in leading order of the $1/N_c$ expansion;
(c) the helicity amplitudes with quark helicities $--$ can be expressed in
terms of those with $++$ by the parity relations Eq.~(\ref{parity}). In this way we obtain
\begin{subequations}
\ba
\label{amp_largen_chiraleven_begin}
 \sigma^0 \tau^0 &\rightarrow& 2( A_{++,++}^{u + d} + A_{-+,-+}^{u + d}) , \\
 \sigma^3 \tau^3 &\rightarrow& 2( A_{++,++}^{u - d} - A_{-+,-+}^{u - d}) , \\
 \sigma^1 \tau^3 &\rightarrow& 2( A_{++,-+}^{u - d} + A_{-+,++}^{u - d}) , \\
i\sigma^2 \tau^3 &\rightarrow& 2( A_{++,-+}^{u - d} - A_{-+,++}^{u - d}) ,
\label{amp_largen_chiraleven_end}
\ea
\end{subequations}
in the sense that the corresponding structures in the large--$N_c$ correlators
Eqs.~(\ref{Eq:YYa}) and (\ref{Eq:YYb}) are to be identified with the given
combination of helicity amplitudes. The content of the relations
Eqs.~(\ref{amp_largen_chiraleven_begin})--(\ref{amp_largen_chiraleven_end})
is identical to that of Eqs.~(\ref{Eq:amplitudes-expand-1})--(\ref{Eq:amplitudes-expand-1-end}),
if one substitutes the large--$N_c$ expressions for the nucleon GPDs in terms
of the soliton GPDs. The chiral-odd correlator Eq.~(\ref{Eq:XXc}) in the same representation
takes the form
\ba\label{Eq:YYc}
{\cal M}(i\sigma^{+1})
&=& \sigma^0 \, \tau^0 \, \delta_t \, \bar{E}_{T, {\rm sol}}
\; - \; \frac{\sigma^2 \, \tau^3}{3} \, ( H_{T, {\rm sol}} + \xi 
\widetilde{E}_{T, {\rm sol}}) ,
\\[1ex]
{\cal M}(i\sigma^{+2})
&=& \frac{\sigma^1 \, \tau^3}{3} ( H_{T, {\rm sol}} + \xi \widetilde{E}_{T, {\rm sol}})
- \frac{\sigma^3 \, \tau^3}{3} \, 2 \delta_t \widetilde{E}_{T, {\rm sol}} .
\ea
We now use that
\ba
i\sigma^{+1} &=& \phantom{-i} 2 (\Gamma_{+-} - \Gamma_{-+}) ,
\\[1ex]
i\sigma^{+2} &=& -2i (\Gamma_{+-} + \Gamma_{-+}) ,
\ea
express the amplitudes with quark helicities $+-$ in terms of those with 
$-+$ using the parity relations Eq.~(\ref{parity}), and obtain
\begin{subequations}
\ba
\label{amp_largen_chiralodd_begin}
 \sigma^0 \tau^0 &\rightarrow& \phantom{-i}2( A_{++,+-}^{u + d} + A_{-+,--}^{u + d}) , \\
 \sigma^2 \tau^3 &\rightarrow& \phantom{-}2i ( A_{++,--}^{u - d} - A_{-+,+-}^{u - d}) , \\
 \sigma^1 \tau^3 &\rightarrow& -2i( A_{++,--}^{u - d} + A_{-+,+-}^{u - d}) , \\
 \sigma^3 \tau^3 &\rightarrow& -2i( A_{++,+-}^{u - d} - A_{-+,--}^{u - d}) ,
\label{amp_largen_chiralodd_end}
\ea
\end{subequations}
to be understood in the same sense as
Eqs.~(\ref{amp_largen_chiraleven_begin})--(\ref{amp_largen_chiraleven_end}).
Again, these relations reproduce 
Eqs.~(\ref{Eq:amplitudes-III})--(\ref{Eq:amplitudes-III-end})
if we substitute the specific large--$N_c$ expressions of the nucleon GPDs in terms
of the soliton GPDs.

Equations~(\ref{amp_largen_chiraleven_begin})--(\ref{amp_largen_chiraleven_end}) exhibit
the degeneracy of the large--$N_c$ correlator noted in Sec.~\ref{Sec-3:large-Nc}:
the spin structures $\sigma^1$ and $\sigma^2$ occur with the same coefficient function
and thus cannot be distinguished in the large--$N_c$ nucleon.
This illustrates again that in leading order of the $1/N_c$ expansion there is no correlation
between the transverse nucleon spin, $\sigma^1$ or $\sigma^2$, and the
transverse momentum transfer, $\bm{\Delta}_T=(\Delta^1, 0)$.
\end{document}